\documentclass{emulateapj}

\shorttitle{A More Fundamental Plane}
\shortauthors{Bolton et al.\ }

\begin{document}
 
\title{A More Fundamental Plane\altaffilmark{1}}

\author{Adam S. Bolton\altaffilmark{2}}
\author{Scott Burles\altaffilmark{3}}
\author{Tommaso Treu\altaffilmark{4}}
\author{L\'{e}on V. E. Koopmans\altaffilmark{5}}
\author{Leonidas A. Moustakas\altaffilmark{6}}

\slugcomment{ApJ Letters, in press}

\altaffiltext{1}{Based on observations made with the
NASA/ESA 
\textsl{Hubble Space Telescope}, obtained
at the Space Telescope Science Institute, which is operated
by AURA, Inc., under NASA contract NAS 5-26555.
These observations are associated with programs \#10174
and \#10587.}
\altaffiltext{2}{Harvard-Smithsonian Center for Astrophysics, 60 Garden
St., Cambridge, MA 02138 ({\tt abolton@cfa.harvard.edu})}
\altaffiltext{3}{Department of Physics and Kavli Institute for
Astrophysics and Space Research, Massachusetts Institute of Technology,
77 Massachusetts Avenue, Cambridge, MA 02139 ({\tt burles@mit.edu})}
\altaffiltext{4}{Department of Physics, University of California,
 Santa Barbara, CA 93101, USA ({\tt tt@physics.ucsb.edu})}
\altaffiltext{5}{Kapteyn
 Astronomical Institute, University of Groningen, P.O. Box 800, 9700AV
 Groningen, The Netherlands ({\tt koopmans@astro.rug.nl})}
\altaffiltext{6}{Jet Propulsion Laboratory, Caltech, MS 169-327, 4800 Oak Grove Drive, Pasadena, CA 91109
({\tt leonidas@jpl.nasa.gov})}

\begin{abstract}
We combine strong-lensing masses with SDSS stellar velocity
dispersions and \textsl{HST}-ACS effective (half-light) radii for 36
lens galaxies from the Sloan Lens ACS (SLACS) Survey to
study the mass dependence of mass-dynamical
structure in early-type galaxies.
We find that over a 180--390\,km\,s$^{-1}$ range in velocity dispersion,
structure is independent of lensing mass to within 5\%.
This result suggests a systematic
variation in the total (i.e., luminous plus dark matter)
mass-to-light ratio as the origin of
the \textit{tilt} of the
fundamental plane (FP) scaling relationship between
galaxy size, velocity dispersion, and surface brightness.
We construct the FP of the lens sample, which we find to be consistent with
the FP of the parent SDSS early-type galaxy population,
and present the first observational correlation between
mass-to-light ratio and residuals about the FP\@.
Finally, we re-formulate the FP in terms of surface
mass density rather than surface brightness.
By removing the complexities of stellar-population effects,
this mass-plane formulation will facilitate comparison to numerical
simulations and possible use as a cosmological distance
indicator.
\end{abstract}


\section{Introduction}

The \textit{fundamental plane} \citep[FP;][]{dr_fp, dd_fp}
is a well-known scaling relationship
between the size $R$, surface brightness $I$,
and velocity dispersion $\sigma$
of elliptical galaxies, expressed in the form
\begin{equation}
R \propto \sigma^a I^b~,
\end{equation}
that indicates an underlying regularity
within the population.  Crudely speaking,
the exponents of the FP are
$(a,b) = (1.5,-0.8)$ (e.g.\ \citealt{bernardi_fp}),
differing significantly from the most na\"{i}ve
constant mass-to-light dimensional-analysis expectation
of $(a,b) = (2,-1)$.  This so-called \textit{tilt}
of the FP (relative to the \textit{virial plane})
can be attributed to either a systematic variation
in the mass-dynamical structure of ellipticals
(\textit{structural non-homology}), a systematic
variation of the stellar mass-to-light ratio,
or a systematic variation in the central dark-matter
fraction with other quantities.
This ambiguity is difficult to resolve, since neither mass
structure (as opposed to light structure), nor mass-to-light
ratio, nor dark-matter fraction
are directly and independently observable
(see e.g.\ \citealt*{clr_fp_96}).
As a result, no definite consensus has emerged as to the
underlying explanation of the FP\@.

Strong gravitational lensing can break some of this
degeneracy observationally, since the total mass
within the Einstein radius $R_{\mathrm{Ein}}$
is directly observable in gravitational lens galaxies
independent of any dynamical modeling.
The main impediment to this strong lensing approach has been the
lack of a large and homogeneous sample of strong lenses for which
the traditional FP observables $R$,
$I$, and $\sigma$ can all be
reliably measured.  This limitation has recently been overcome
by the Sloan Lens ACS (SLACS) Survey (\citealt{slacs1, slacs2, slacs3,
slacs4}; hereafter Papers I--IV
respectively; see also \citealt{bolton_speclens, bolton_1402}),
an ongoing survey
for strong gravitational lens galaxies combining spectroscopic
lens-candidate selection from the Sloan Digital Sky
Survey (SDSS; \citealt{york_sdss}) with high-resolution follow-up imaging
with the Advanced Camera
for Surveys (ACS) aboard the \textsl{Hubble Space Telescope}
(\textsl{HST}).
In this \emph{Letter}, we investigate
the dependence upon mass of early-type galaxy
mass-dynamical structure by adding strong-lensing
information to the traditional FP observables.
We also construct the FP itself for the SLACS
sample, show that the residuals about the best fitting FP are significantly
correlated with mass-to-light ratio, and present a
new formulation of the
FP using lensing data to replace surface brightness with
surface mass density.

\section{Observations and Measurements}

The imaging data for all analysis
presented in this \emph{Letter} were collected with single
420-s exposures through the F814W ($I$-band) filter
under \textsl{HST} programs
10174 (PI: Koopmans) and 10587 (PI: Bolton).
As of 2006 July,
89 systems were successfully observed with ACS-WFC
by these two Snapshot programs, yielding 44
strong lenses with redshifts in the
range $z_{\mathrm{lens}} \simeq 0.1$--0.4.  Full details
of the final SLACS Snapshot sample will be presented in
a forthcoming paper (Bolton et al.\ 2007, in preparation;
hereafter B07).
To generate lens-galaxy-subtracted images suitable
for gravitational lens modeling and to measure lens-galaxy
flux within an arbitrary radial aperture, we
fit a PSF-convolved elliptical
b-spline model to the surface brightness profile of
the lensing galaxies (see Paper~I).
We also fit a PSF-convolved
elliptical deVaucouleurs surface brightness
model to the image of each lens galaxy, to measure
intermediate-axis half-light (effective) radii ($R_e$)
and brightnesses ($I_e = 0.5 L / \pi R_e^2$)
for use in the FP analysis.  Lastly, we make lensing mass
measurements by fitting
singular isothermal ellipsoid (SIE; e.g.\ \citealt{kormann_sie})
mass models to the
extended multiple images of lensed background galaxies.
We correct the \textsl{HST} $I$-band photometry with
the Galactic dust corrections of \citet*{sfd_dust},
synthetic $k$-corrections to rest-frame
$V$-band based on SDSS spectral templates
for each galaxy, an evolution correction of
$d \log \Upsilon_V / dz = -0.4$
\citep{kelson_2000_iii, treu_2001_iii, moran_2005_iii}, and
distance moduli computed for an
$(\Omega_M, \Omega_{\Lambda},h) = (0.3,0.7,0.7)$ FRW universe.
Throughout this \emph{Letter}, the aforementioned FRW
cosmology is assumed, magnitudes are in the AB system,
masses and luminosities are in solar units,
sizes are expressed in kpc, velocities are in km\,s$^{-1}$,
and logarithms are base 10.

In order to achieve a uniform physical
aperture with as little extrapolation and artificial
covariance as possible, we correct all lensing mass
measurements to an aperture of $R_e / 2$ using the isothermal mass model,
since the median ratio of $R_{\mathrm{Ein}}$
to $R_e$ within our lens sample is 0.55.
The isothermal model is preferred by the
lens-dynamical analysis of Paper III,
and is also favored by statistical studies of
lensed quasars \citep[e.g.][]{rusin_kochanek_05}
and by dynamical analysis of nearby ellipticals
\citep[e.g.][]{gerhard_2001}.
To assess the sensitivity of our results to the
isothermal assumption, we derive alternate
aperture-mass corrections using a light-traces-mass (LTM)
model determined from the b-spline photometry.
The mean difference between isothermal and LTM
aperture masses within $R_e / 2$ is $-3$\%,
with an RMS difference of $\pm 9$\%.
To compute central mass-to-light ratios
(which we denote by $\Upsilon_{e2}$), luminosities within
$R_e / 2$ are measured from the b-spline models.
We also correct the observed velocity
dispersions from the SDSS fiber
radius of $1\farcs5$ to $R_e / 2$ (giving $\sigma_{e2}$)
using the empirical
relation of \citet*{jorgensen_vdisp}.  These velocity
corrections are small: 1.4\% $\pm$ 1.3\%.

We reject 8 lenses based upon (i) prominent
spiral arms, (ii) an SDSS spectral
SNR per pixel of less than 7, or (iii)
significant disagreement between the
model $R_e$ and the radius of an aperture
containing one-half of the model flux.
These cuts leave
us with a sample of 36 early-type strong lens galaxies
with absolute evolution-corrected magnitudes $M_V$ in the
range $-20.8$ to $-24.0$ and velocity dispersions
$\sigma_{e2} \simeq 180$--390\,km\,s$^{-1}$.
Quoted errors on all fitted parameters
are 68\% limits from bootstrap re-sampling
of these 36 lenses.

\section{The Mass (In)Dependence of Structure}
\label{htest}

Dimensional analysis of the Jeans equation
(e.g., \citealt{bt87}) shows that
\begin{equation}
M = c \sigma^2 R / G~,
\end{equation}
where $R$ is a scaling radius, $\sigma$ is an average
stellar velocity dispersion within that radius, $M$ is the
mass interior to that radius, $G$ is Newton's constant,
and $c$ is a dimensionless structure
constant.  Note that $R$ and $\sigma$
pertain to the luminous (i.e., stellar) component, while $M$
refers to the \textit{total} mass (i.e.\ stellar and dark):
this follows from
the Jeans equation, and matches well with the quantities
that we actually measure with photometry, spectroscopy,
and lens modeling.
The value of $c$ will in general depend upon the anisotropy
of the stellar orbits, the overall shape of both the luminous- and
dark-matter density profiles, and the relative fraction
of luminous to dark matter in the region of interest.
In the SLACS sample, a systematic variation of $c$
with galaxy mass will be apparent in
the relationship between the lensing-determined total mass
$M_{\mathrm{lens}}$ within $R_e / 2$ on the
one hand and the dimensional mass variable
$M_{\mathrm{dim}} = G^{-1} \sigma_{e2}^2 (R_e/2)$ on the
other, which we parameterize as
\begin{equation}
\log M_{\mathrm{lens}} = \delta \log M_{\mathrm{dim}} + \log c_0~.
\end{equation}
A systematic trend of $c$ with mass
thus corresponds to a value of $\delta$ different than 1.
Minimizing the scatter orthogonal to the best-fit relation,
we find $\delta = 0.986 \pm 0.034$ ($\delta = 0.956 \pm 0.042$)
and $\log c_0 = 0.58 \pm 0.16$ ($\log c_0 = 0.74 \pm 0.19$)
for isothermal (LTM) aperture mass corrections.
This relationship is shown in the
left-hand panel of Fig.~\ref{fp_ml_3}.
We see that---whatever the underlying details---the SLACS lens
population is empirically consistent with \textit{no} variation of
the structure constant $c$ with mass.
This result is closely related to the nearly
one-to-one relation between stellar and isothermal
lens-model velocity dispersions
shown for the initial SLACS sample in \citet{slacs2}.
By casting this result in terms of nearly model-independent
aperture masses,
the current analysis emphasizes the strong evidence for
the mass independence of early-type galaxy structure
in the range $\sigma_{e2}=$180--390\,km\,s$^{-1}$,
\textit{independent of any
assumed underlying mass-dynamical model}.

\section{Planes Fundamental}

\begin{figure*}[t]
\plotone{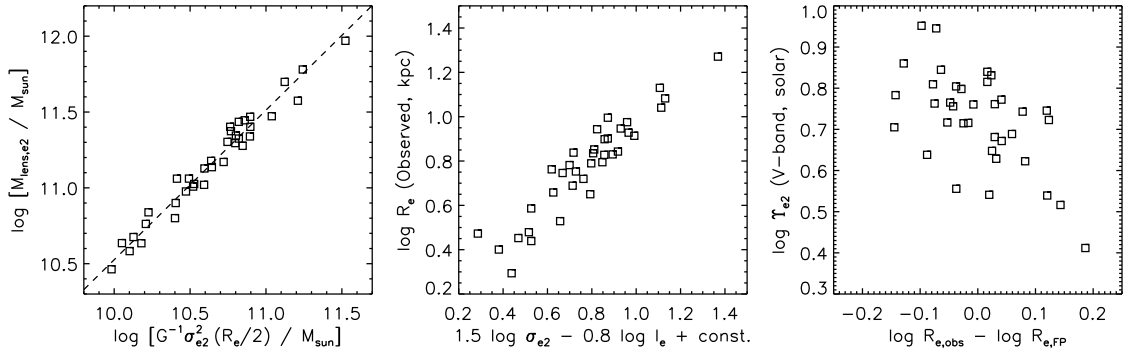}
\caption{\label{fp_ml_3}
\textsl{Left}: relationship between lensing mass and dimensional
mass within $R_e / 2$.  Dashed line is best fitting linear
relation between these logarithmic quantities,
with slope of $0.986 \pm 0.034$.
\textsl{Center}: effective radius $R_e$ as observed and as
predicted from the best-fitting FP relationship between
$R_e$, $I_e$, and $\sigma_{e2}$.  \textsl{Right}: mass-to-light
ratio $\Upsilon_{e2}$ in the $V$-band versus residuals about the
FP relationship.}
\end{figure*}

To place the above result in the context of the FP,
we must first verify that the SLACS sample does indeed define the
same FP as non-lenses.
Paper II showed the original SLACS sample of 15 lenses
to be consistent with the locally determined FP when
corrected for luminosity evolution; the current, larger sample
allows us to directly constrain the FP of the lenses.
We perform this fit in the orthogonal sense,
defining the best-fitting plane as that which minimizes
the total squared perpendicular distance from all data
points in the space spanned by $\log R_e$, $\log I_e$,
and $\log \sigma_{e2}$.  Defined in this way, the coefficients
of the FP do not change with the choice of dependent variable.
Expressing the FP in the form
\begin{equation}
\label{fpdef}
\log R_e = a \log \sigma_{e2} + b \log I_e + d~,
\end{equation}
we find $a = 1.50 \pm 0.32$, $b = -0.78 \pm 0.13$,
and $d = 3.9 \pm 1.7$.  The given errors are the
square-root diagonal entries of the covariance matrix of
a set of bootstrap-re-sampled coefficient fits; off-diagonal
correlations are
$\rho_{ab} \simeq 0.6$, $\rho_{ad} \simeq -0.8$,
and $\rho_{bd} \simeq -0.9$.
The residual logarithmic orthogonal
scatter about the best-fit plane is 0.041\,dex.
The FP is shown in edge-on projection with respect
to $R_e$ in the center panel of Fig.~\ref{fp_ml_3}.
The scaling coefficients ($a$ and $b$)
of the FP defined by the SLACS
lens sample are consistent with
the orthogonal FP fits of \citet{bernardi_fp} for
early-type galaxies from the SDSS (with somewhat different
selection and conventions).

As an aside, the addition of lensing data allows us to make the
first direct observation of a correlation between
the residual scatter about the FP and the total
mass-to-light ratio $\Upsilon_{e2}$ as determined from lensing,
shown in the right-hand panel of Fig.~\ref{fp_ml_3}.
The correlation has a linear (Pearson) coefficient of $-0.58$, giving
a formal significance of 99.98\% for a sample size of 36.  The significance
of the correlation is similar under the LTM assumption (i.e.\ no correction
of $\Upsilon$ from $R_{\mathrm{Ein}}$ to $R_e / 2$).
Furthermore, the FP residuals do not have any significant correlation
with mass or with luminosity separately, nor any
correlation with redshift or with the ratio $R_{\mathrm{Ein}} / R_e$.
We are thus drawn to the conclusion that this
correlation is intrinsic, and not introduced through a
correlation of measurement errors.
Though they did not observe it directly, such a correlation
of FP residuals with mass-to-light ratio
was deduced by \citealt{faber_87} to be the most
likely explanation for the intrinsic thickness of the FP\@.
(Also see \citealt*{jorgensen_fp} for an extensive observational
analysis of FP residual correlations and intrinsic thickness.)

We now use the strong-lensing data to
formulate an analogous plane in mass space,
which we will refer to as the \textit{mass plane} (MP)\@.
Following the traditional form
of the FP, we define the MP by replacing surface brightness $I_e$ with
surface mass density within $R_e / 2$, denoted by $\Sigma_{e2}$
(in $M_{\sun}$\,kpc$^{-2}$):
\begin{equation}
\log R_e = a_m \log \sigma_{e2} + b_m \log \Sigma_{e2} + d_m~.
\end{equation}
Note that in going from $I_e$ to $\Sigma_{e2}$,
the sensitivity to luminosity evolution with redshift is removed.
Fitting as for the FP above, the MP coefficients are
$a_m = 1.77 \pm 0.14$, $b_m = -1.16 \pm 0.09$, and $d_m = 7.8 \pm 1.0$.
Off-diagonal correlations are $\rho_{ab,m} \simeq 0.1$,
$\rho_{ad,m} \simeq -0.4$, and $\rho_{bd,m} \simeq -0.9$.  The RMS orthogonal
logarithmic scatter about the MP is 0.026\,dex.  Fitting for the MP
with mass densities computed using the LTM assumption
rather than the isothermal model gives $a_m = 1.86 \pm 0.17$,
$b_m = -0.93 \pm 0.09$, and $d_m = 5.4 \pm 0.9$, with comparable
parameter correlations and an orthogonal logarithmic
scatter of 0.030\,dex.  The MP thus appears to be \textit{tighter} than
the FP in the sense of having smaller residual scatter.
This scatter is consistent with the $\sim$7\% fractional
velocity-dispersion errors in the current data.

The significant
difference between the FP and the MP is most easily seen by
considering the parameterization of the FP tilt in terms of a
relationship of the form $L \propto M_{\mathrm{dim}}^{\eta}$
(e.g.\ \citealt{dr_fp, clr_fp_96}, \citealt*{tbb_fp}, \citealt{treu05b}).
While this form does not exactly reproduce
the best-fitting FP exponents of the current study, it is
a convenient physically motivated approximation.
Fitting in log-space by minimizing the orthogonal
scatter, we find $\eta = 0.82 \pm 0.05$ for the SLACS sample.
This translates into FP coefficients of $a = 1.39$ and $b = -0.85$
in the form of Eq.~\ref{fpdef}.
Comparing this with the result that $\delta = 0.986 \pm 0.034$
from \S~\ref{htest}---which amounts to an
analogous parameterization of the MP---we see that in
going from the FP to the MP, the tilt relative to the virial
relation is essentially eliminated.

\section{Discussion}

Previous studies have constructed the FP of gravitational lens
galaxies \citep{kochanek_lens_fp, vandeven_fp, rusin_kochanek_05},
but have taken their velocity dispersions from the strong
lensing data rather than from stellar dynamics.  By combining
\textit{independent} measurements of mass and velocity,
we have found direct evidence that the total mass-dynamical structure
of early-type galaxies is independent of mass.
What, then, does this result tell us about the underlying
explanation for the tilt of the FP?  The straightforward
binary answer is that
the FP must be due to a systematic variation in the
total central mass-to-light ratio---either through the
stellar population or through the central dark-matter
fraction---and not due to any
systematic variation of mass-dynamical structure
with mass.
If mass-dynamical structure effects were responsible
for the FP, they should also manifest as a systematic variation
with mass of the structure constant $c$ of \S~\ref{htest}
and a similar tilt in the MP as in the FP, whereas we
see neither of these two effects.
This conclusion is most
notably consistent with those of
\citet{padmanabhan_04} and \citet{capp_sauron};
the agreement is reassuring given the diversity of
the methods employed between these studies and ours.

Our conclusions are at variance
with those of \citet{tbb_fp}, who point to
non-homology in the \textit{luminous} component of early-type galaxies
(specifically, the systematic variation
in S\'{e}rsic index with galaxy luminosity) as being
largely responsible for the FP\@.
A plausible explanation is that,
motivated by the results of \citet{romanowsky_pne},
\citet{tbb_fp} do not allow for the presence of dark matter.
Other lines of evidence suggest a ``bulge-halo conspiracy''
in early-type galaxies, whereby the combined luminous and dark-matter
profiles generate a more uniformly flat
rotation curve than either component alone
(e.g.\ \citealt{gerhard_2001}).
Such an effect could offset non-homology
in the luminous component, leading us
back to systematic mass-to-light variation as the explanation for the FP\@.

As mentioned above, our
observations do not distinguish between stellar mass and dark mass:
the two physically very distinct possibilities
of a variation in the stellar mass-to-light ratio and a variation in
the central dark-matter fraction are (in principle) equally allowed.
Previous studies have shown that variations
in stellar mass-to-light
must be responsible for some (but not all) of the tilt
of the FP, as seen through the systematic decrease in tilt at
redder wave-bands \citep{pahre_iv}, detailed
stellar population modeling \citep{thomas_05},
and the galaxy-mass-dependent average rate of
luminosity evolution with redshift \citep{treu05a, treu05b}.
At this stage, the main diagnostic available to us to quantify stellar
populations effects in the SLACS sample is the strength of the
4000\AA\ break in lens-galaxy continuum flux $D_n(4000)$, which should
correlate with stellar mass-to-light ratio (\citealt{bruzual_83,
hamilton_85, balogh_99, kauffmann_03}). No significant correlation
with $\Upsilon_{e2}$ is found, although the sensitivity of $D_n(4000)$
may not be sufficient to detect the small age differences ($\Delta t/ t
\sim 2/10$) expected for early-type galaxies in this range of velocity
dispersion \citep{thomas_05}. Taken at face value, the result suggests
that most of the variation in $\Upsilon_{e2}$ is due to dark matter
content.  We will quantify this statement in future work with
the aid of more sophisticated stellar population diagnostics.

In the process of investigating the mass dependence
of early-type galaxy structure and the genesis
of the FP, we have made the first observational
correlation of the thickness of the FP with
mass-to-light ratio, and
we have introduced an analogous plane in mass space that is both tighter
and less tilted than the FP relative to the virial scaling relation.
This observational mass plane
is a fundamental dynamical scaling relation for early-type
galaxies; though it requires a measurement of galaxy \textit{size},
it makes no reference to galaxy \textit{luminosity}, and therefore
it can be tracked across cosmic time without regard to stellar evolution.
The MP thus offers great promise as a cosmological standard ruler,
particularly if the relation can be calibrated
locally by masses from detailed dynamical models of nearby
elliptical galaxies.  (Though structural evolution may still occur:
e.g.\ \citealt{vdm_vd_06a, vdm_vd_06b}.)
By removing the need for stellar-population
modeling, the MP is also more suitable for comparison with
the results of numerical simulations.

This \emph{Letter} has not addressed possible biases introduced
through our gravitational-lens selection procedure.
We defer a full discussion
of selection effects to B07, but note that
Papers I and II showed the original SLACS sample to
be statistically consistent with the parent sample of
non-lens SDSS galaxies from which they were selected.

The analysis presented here has
been made possible only recently
by the SLACS Survey gravitational lens sample.
Existing and forthcoming deep \textsl{HST}
imaging will permit more accurate mass,
luminosity, and color measurements,
while ongoing spectroscopy with the VLT and Keck telescopes
will provide more robust velocity dispersion measurements
and stellar-population diagnostics.
These data will soon enable a greater precision
in the type of analysis described in this \emph{Letter}.

\acknowledgements
The authors thank G. Bertin, L. Ciotti, A. Renzini,
and the anonymous referee for valuable comments.
ASB, TT, LVEK, and LAM acknowledge the support
and hospitality of the Kavli
Institute for Theoretical Physics
at UCSB, where a significant part of
this work was completed.  This research was
supported in part by the NSF under
Grant No.\ PHY99-07949.
TT acknowledges support from the NSF through
CAREER award NSF-0642621, and from the Sloan Foundation
through a Sloan Research Fellowship.
LVEK is supported
in part through an NWO-VIDI program subsidy (project \#639.042.505).
The work of LAM was carried out at JPL/Caltech,
under a contract with NASA\@.
Support for \textsl{HST} programs \#10174 and \#10587 was provided
by NASA through a grant from STScI, which is operated by AURA, Inc.,
under NASA contract NAS 5-26555.  Funding for the SDSS has been provided by
the Alfred P. Sloan Foundation, the Participating Institutions,
the NSF, the U.S. D.O.E.,
NASA, the Japanese Monbukagakusho, the Max Planck Society, and
the Higher Education Funding Council for England.
The SDSS Web Site is \url{http://www.sdss.org/}.
The SDSS is managed by the Astrophysical Research
Consortium for the Participating Institutions.


\end{document}